\documentclass[aps,prl,twocolumn,showpacs,superscriptaddress,groupedaddress,nofootinbib]{revtex4}
\usepackage[utf8x]{inputenc}
\usepackage{graphicx}
\usepackage{amsmath,amssymb} 
\usepackage[english]{babel}
\usepackage{rotating}
\usepackage{amsbsy}
\usepackage{textcomp}
\usepackage{psfrag}
\usepackage{multirow}
\usepackage{color} 
\usepackage{sidecap}
\usepackage{array}
\usepackage{xspace}
\usepackage{subfig}
\usepackage{natbib}
\usepackage[breaklinks,colorlinks,citecolor=blue]{hyperref}

\newcommand{\beq }{\begin{equation}}
\newcommand{\eeq}{ \end{equation}}
\newcommand{\beqa }{\begin{eqnarray}}
\newcommand{\eeqa }{\end{eqnarray}}
\newcommand{\bwt }{\begin{widetext}}
\newcommand{\ewt }{\end{widetext}}
\newcommand{\bef}{\begin{figure}[htb!]}
\newcommand{\eef}{\end{figure}}

\newcommand{\apjl}{\emph{Astrophys. J. Letters \xspace}}

\newcommand{\gws}{GWs\xspace}
\newcommand{\gw}{GW\xspace}

\newcommand{\msun}{$M_\odot$\xspace}

\newcommand{\w}{$f_{a}$\xspace}

\usepackage{aas_macros}
\newcommand{\Ha}{$\mathcal{H}_a$\xspace}
\newcommand{\Hi}{$\mathcal{H}_{\bar{a}}$\xspace}
\newcommand{\si}{$\sim$\xspace}

\begin{document}

\title{Use of gravitational waves to probe the formation channels of compact binaries}
\author{Salvatore Vitale}
\email[email:  ]{salvatore.vitale@ligo.org}
\author{Ryan Lynch}
\affiliation{Massachusetts Institute of Technology, 185 Albany St, 02138 Cambridge USA}
\author{Riccardo Sturani}
\affiliation{ICTP South American Institute for Fundamental Research\\
Instituto de F\'\i sica Te\'orica, Universidade Estadual Paulista,
  S\~ao Paulo, SP 011040-070, Brasil}
\author{Philip Graff}
\affiliation{Department of Physics \& Joint Space-Science Institute, University of Maryland, College Park, MD 20742, USA}
\affiliation{Gravitational Astrophysics Laboratory, NASA Goddard Space Flight Center, 8800 Greenbelt Rd., Greenbelt, MD 20771, USA}
\begin{abstract}

With the discovery of the binary black hole coalescence GW150914, the era of gravitational-wave astrophysics has started.
Gravitational-wave signals emitted by compact binary coalescences will be detected in large number by LIGO and Virgo in the coming months and years. Much about compact binaries is still uncertain, including some key details about their formation channels. 
The two scenarios which are typically considered, common envelope evolution and dynamical capture, result in different distributions for the orientation of the black hole spins. In particular, common envelope evolution is expected to be highly efficient in aligning spins with the orbital angular momentum. 
In this paper we simulate catalogs of gravitational-wave signals in which a given fraction of events comes from common envelop evolution, and has spins nearly aligned with the orbital angular momentum. We show how the fraction of aligned systems can be accurately estimated using Bayesian parameter estimation, with 1 $\sigma$ uncertainties of the order of 10\% after 100-200 sources are detected. 
\end{abstract}
\maketitle

\section{Introduction}

Advanced LIGO~\cite{TheLIGOScientific:2014jea} has just completed its first science run. Results from the first 5 weeks of data have been made public, and include the binary black hole (BBH) coalescence GW150914~\cite{GW150914-DETECTION}.  Advanced Virgo~\cite{TheVirgo:2014hva} is expected to join LIGO in the second science run~\cite{2016LRR....19....1A}, starting in Fall 2016. KAGRA~\cite{PhysRevD.88.043007} and LIGO India~\cite{M1100296} are expected to join the network before the end of this decade. 
Ground based interferometers will detect gravitational radiation from several kinds of sources and start gravitational-wave (\gw) astrophysics.
Compact binary coalescences (CBC) of two neutron stars (BNS), two black holes or a neutron star and a black hole (NSBH) have traditionally been among the most promising sources, and will be detected at a rate of several tens per year, although significant uncertainty still exists in the astrophysical rates, both predicted and measured~\cite{GW150914-RATES,2010CQGra..27q3001A}. Analysis of many such signals promises to shed light on several open problems in astrophysics. 
For example, direct mass measurement with \gws could allow for an accurate reconstruction of black hole and neutron star mass functions. \gws also represent our first chance to perform strong field tests of general relativity. An idea of what can be done is given by the recent analyses of GW150914~\cite{GW150914-PARAMESTIM,GW150914-TESTOFGR,GW150914-ASTRO}. In the last few years a pipeline has been created, called TIGER~\cite{2012PhRvD..85h2003L,2012JPhCS.363a2028L,2014PhRvD..89h2001A}, which can look for unmodeled deviations from general relativity in detected \gw signals. Elsewhere~\cite{2013PhRvL.111g1101D} it has been shown how \gws could be used to test proposed equations of state for neutron stars. Most of this work, and others~\cite{2012PhRvD..86d4023L,2014arXiv1410.3852L}, relies in Bayesian model selection, instead of simple parameter estimation. This has the advantage that information can be aggregated from all detected signals, in a cumulative way, resulting in more powerful tests.
In this letter we show that \gws can be used to check whether spins are preferentially aligned with the orbital angular momentum in BBH and NSBH systems (we will not consider BNS, since known neutron stars in binaries do not have large spins~\cite{2003Natur.426..531B}). This is of fundamental importance for astrophysics and for understanding the formation mechanisms of compact binaries, for which there still are many open questions~\cite{2010ApJ...725.1984L,2015MNRAS.447.2181I,2014arXiv1405.4662Z,2005ApJ...632.1035O,2016PhRvD..93h4029R,2016MNRAS.458.2634M,2016arXiv160404254R}. 
It is believed that two main formation channels exist for compact binaries (See~\cite{GW150914-ASTRO} for a review). Common envelope evolution is expect to happen in galactic fields, whereas dynamical capture could happen in dense environments such as globular clusters. Critically, it has been suggested that common envelope evolution in binaries will align the spins with the orbital angular momentum~\cite{2007ApJ...661L.147B}~\footnote{Others suggest that kicks introduced in the system when the progenitor stars undergo core collapse supernovae could result in spins being significantly misaligned~\cite{2000ApJ...541..319K}}. Spins are instead expected to be randomly oriented for CBCs formed dynamically. Ultimately, being able to verify if and how often spins are aligned could significantly help to understand the formation patterns of binary systems, verify which channel happens more frequently, and the efficiency of the common envelope evolution phase in aligning spins.
In this paper we consider a scenario where a fraction \w of signals have spins nearly aligned, while the rest are non-aligned. This accounts for two possible formation patterns for CBCs, one of which is efficient in aligning spins with the orbital angular momentum. We find that the posterior distribution for the mixture parameter \w can be accurately estimated with a couple hundred sources, with precision of the order of $10\%$.

\section{Method}

Let us assume two main formation channels for CBC exist, which result in a fraction \w of systems having spins nearly aligned, and a fraction (1-\w) having misaligned spins. We will assume $N$ \gw detections are made, denoted by their data streams $d_i\mbox{ with } i=1 \ldots N$, and show how they can be used to estimate \w. 

We introduce two mutually exclusive models: \Ha corresponding to spins nearly aligned with the orbital angular momentum, and \Hi corresponding to non-aligned spins (we will define these models more precisely later).
Given an event and the corresponding data stream $d_k$, we can calculate the evidence of the data for the models above,  $Z^m_k\equiv p(d_k|\mathcal{H}_m)$, with $m=a,\bar{a}$.  The evidence must be calculated by integrating a non-trivial likelihood function over a multidimensional parameter space~\cite{2014arXiv1409.7215V}. Calling $\vec\theta$ the unknown parameters on which a CBC depends, we have:
    
\beq\label{eq.evidence}
Z^m_k = \int_{\Theta_m}{ p(d_k| \vec\theta\, \mathcal{H}_m)p_m(\vec\theta|\mathcal{H}_m)}d\vec\theta \:.
\eeq

We solve this integral by using the Nested Sampling and the BAMBI flavors of \texttt{lalinference}~\cite{2014arXiv1409.7215V}.
We stress that the hypervolume we integrate over, $\Theta_m$, depends on the model being considered (see next section). 

We can now show how the aligned fraction of events can be calculated. 
We start by applying Bayes' theorem and the product rule to the posterior distribution of \w:

\beq
p(f_{a}|\vec{d} )\propto p(\vec{d}|f_{a} )p(f_{a})=p(f_{a})\prod_{k=1}^{N}p(d_k|f_{a} )\label{Eq.PostMixture}
\eeq
where $p(f_{a})$ is the prior on the mixture parameter, that we take as flat on $[0,1]$ since we do not have any previous astrophysical information. 
With $\vec{d}$ we have denoted the set of N detected events, $\vec{d}\equiv \{d_1, d_2, \cdots, d_N\}$.
The factors inside the product can be expanded by noticing that \Ha and \Hi are mutually exclusive for each detection, i.e. that $p(\mathcal{H}_a) + p(\mathcal{H}_{\bar{a}})=1$.
Thus:
    
\beq
p(d_k|f_{a})=\sum_{j=a,\bar{a}}{p(d_k|\mathcal{H}_j f_{a})p(\mathcal{H}_j|f_{a})}
\eeq

We notice that $\mathcal{H}_j f_{a}$ can be written simply as $\mathcal{H}_j$, since knowing if a signal was aligned or not makes knowing \w irrelevant.
Next, we need to calculate $p(\mathcal{H}_a|f_{a})$ and $p(\mathcal{H}_{\bar{a}}|f_{a})$. These are trivially $p(\mathcal{H}_a|f_{a})=f_a$ and $p(\mathcal{H}_{\bar{a}}|f_{a})=(1-f_a)$: if a fraction $f_a$ of events is aligned, the probability that the aligned model applies to any event, before looking at the data, is $f_a$.

Modulo a normalization constant, the log of Eq.~\ref{Eq.PostMixture} then reads:
    
\beqa
\log p(f_{a}|\vec{d})&=&\log(p(f_a))\label{Eq.logpw}\\
    &+&\sum_{k=1}^{N}{\left(\log Z^a_k +\log\left[f_{a}+ (1-f_{a}) \frac{Z^{\bar{a}}_k}{Z^a_k}\right]\right)}\nonumber
\eeqa

\section{Implementation}\label{Sec.Implementation}

In this section we describe the parameters of the sources we simulate. 
We consider \gws emitted by BBH and NSBH, and for each type of source we generate two catalogs of \gw signals, one corresponding to nearly aligned spins and the other to non-aligned spins.

For the BBH, we use the so called IMRphenomPv2 waveform approximant~\cite{Hannam:2013oca,Khan:2015jqa} (this is one of the two families used for the analysis of GW150914~\cite{GW150914-PARAMESTIM}). For the lighter NSBH we use the SpinTaylorT4 approximant~\cite{2003PhRvD..67j4025B,2006PhRvD..74b9904B}. Unlike IMRphenomPv2, SpinTaylorT4 is an inspiral-only approximant and thus cannot model the merger and ringdown phase of CBCs. However, since the frequency at which the merger happens is roughly inversely proportional to the total mass of the system, merger and ringdown can be neglected as long as the total mass is below $\sim 20M_\odot$~\cite{2014arXiv1404.2382M}, for reasonable signal-to-noise ratios (SNRs). All the NSBH we simulate have total mass below 13\msun. 
In both cases, we work at the highest known post-Newtonian phase order, while neglecting higher-order amplitude corrections. We also neglect tidal contributions in neutron stars~\cite{Lattimer:2013,2014PhRvD..89b1303Y,2014PhRvD..89j3012W,2013PhRvL.111g1101D}.  Both these limitations are due to computational considerations and will not impact our main result.  

For the BBH, we choose to consider heavy black holes of a few tens of solar masses, which we know will be detected in large number by ground based detectors in the coming months~\cite{GW150914-RATES}.  We thus generate component masses uniformly from the range $[30,50]$\msun (in the source frame). The dimensionless spins, $a_i\equiv \frac{c|\vec{ S}_i|}{G m_i^2}$, are uniformly generated in the range $[0,0.98]$, compatible with the range of validity of the waveform approximant~\cite{Hannam:2013oca,Khan:2015jqa}. 
For the NSBH signals, BH masses are in the range $[6,11]$\msun, and NS masses in the range $[1.2,2]$\msun. Black hole spins are uniform in the range $[0,1]$ while for the neutron stars we restrict possible spins to $[0,0.3]$ (the largest measured spin of a NS in a binary system is $0.02$~\cite{2008LRR....11....8L}).
For the NSBH sources, our choice for the mass range of BH is mainly driven by the use of inspiral-only waveforms (IMRphenomPv2 waveforms are not considered reliable for mass ratios above $4-5$~\cite{Hannam:2013oca,Khan:2015jqa}).

The distances are uniform in comoving volume, with a lower network SNR (that is, the root-sum of squares of the SNR in each instrument) cut at $8\sqrt{3}\sim13.9$ for NSBH and $7\sqrt{3}\sim12.1$ for BBH. These correspond to distances up to \si1.2~Gpc for NSBH and \si12~Gpc for BBH. For both BBH and NSBH, the sky position and orientation-polarization of the systems are uniform on the unit sphere.

To verify that the test we propose is self-consistent and does not rely on the exact definition of ``aligned'', we define it in a different way for NSBH and BBH.

For NSBH, the nearly aligned (henceforth just aligned) catalog is made of signals with tilt angles (i.e. the angles between spin vectors and orbital angular momentum) isotropic in the interval $[0,10]^\circ$, i.e. close to the positive direction of the orbital angular momentum.
For the BBH, the tilts in the aligned catalog are in the range $[0,10]^\circ \cup [170-180]^\circ$, i.e. the spin vectors can be along both the positive and negative direction of the orbital angular momentum. For both BBH and NSBH, the non-aligned model is the logical negation of the corresponding aligned model. For example, for NSBH tilts were isotropic in the range $[10,180]^\circ$. The priors on the tilt angles for the \Ha and \Hi models, eq.~\ref{eq.evidence}, are isotropic with cuts that match these intervals.

Each event is added into simulated Gaussian noise corresponding to the design sensitivity of the two Advanced LIGO detectors and the Advanced Virgo detector~\cite{2016LRR....19....1A}.

We analyze all events in the catalogs twice, once with a prior that matches the $H_a$ model, and once with a prior that matches \Hi. These runs provide the evidences of eq.~\ref{eq.evidence} that we can use in eq.~\ref{Eq.logpw} to calculate posterior of the mixture fraction.

\section{Results}
\begin{figure}[htb]
\includegraphics[width=0.5\textwidth]{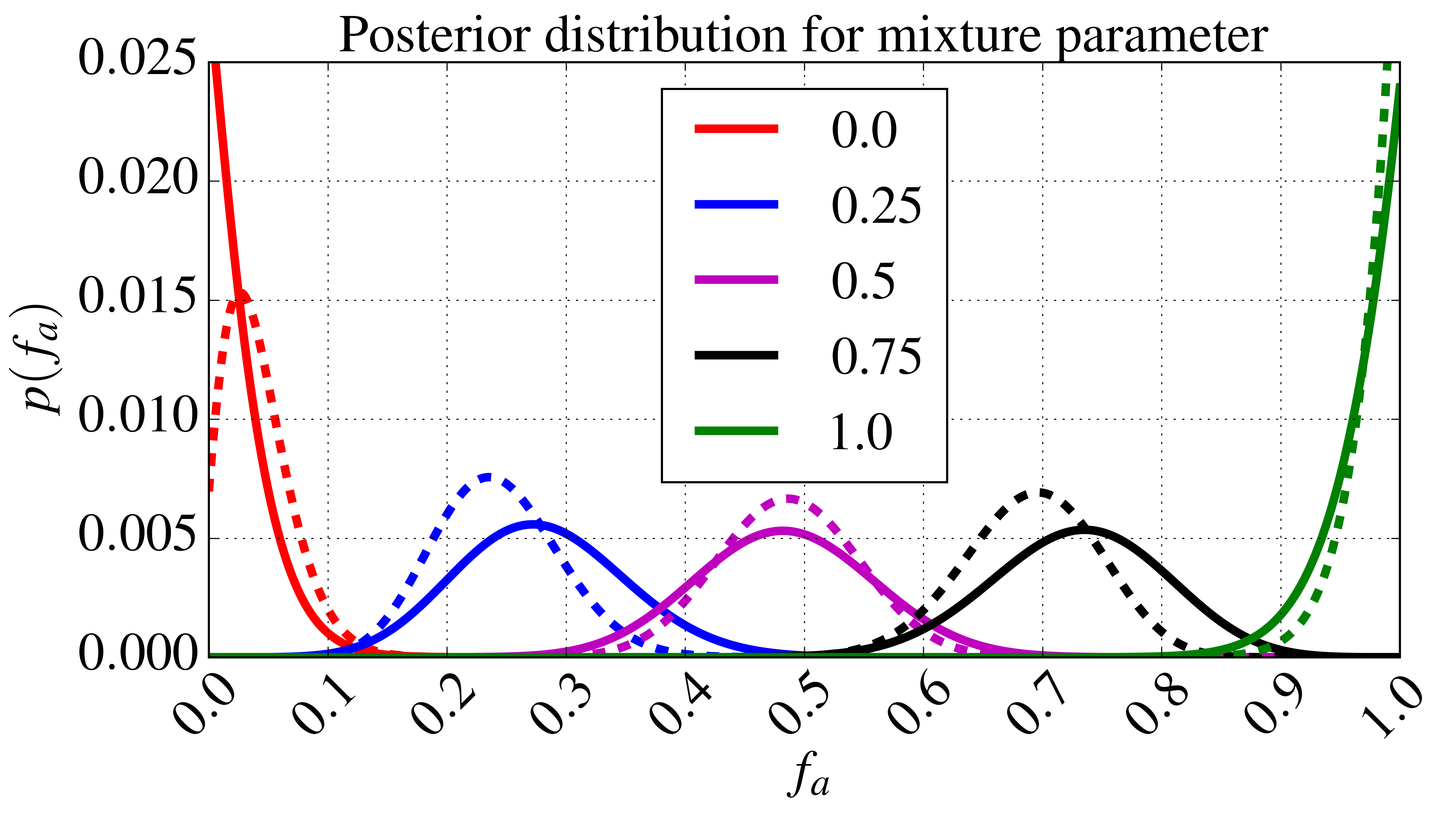}
\caption{Posterior distribution for the mixture parameter \w after 100 NSBH (dashed) and 200 BBH (solid) detections. Several underlying values of \w (given in the legend) are considered. $f_a=0$ corresponds to a catalog where none of the sources had aligned spins, while $f_a=1$ refers to a catalog where all events had aligned spins.}
\label{Fig.Mixture}
\end{figure}

To show how the method performs for some representative values of $f_a$, we generate (for both BBH and NSBH) 5 catalogs with increasing fraction of aligned events. From 0 (all events are not-aligned) to 1 (all events are aligned), with steps of 0.25.
These catalogs are trivially created from our initial set of NSBH and BBH by randomly drawing aligned and non-aligned signals with the desired ratio until 100 sources for NSBH or 200 for BBH are obtained. The evidences of these events are then used in eq.~\ref{Eq.logpw} to obtain the posterior distribution of \w.
The main results are shown in Fig.~\ref{Fig.Mixture}, where for pedagogical purposes we keep separated BBH and NSBH sources.
We see how the posterior distributions for $f_a$ peak at or very close to the corresponding true values, given in the legend, with 1 $\sigma$ uncertainties of the order of $10\%$. The small offsets of some of the curves can be explained with the limited number of events we consider (the offset would be zero in the limit of an infinite number of sources).
By regenerating the catalogs a few times we saw that the peaks can shift by a few percents on either side of the true values. 
We have verified that halving the number of sources in the catalogs (50 NSBH and 100 BBH) broadens the posterior distributions, while leaving them centered around or close to the true values.
One might be surprised that the NSBH distributions are narrower in spite of the fact that fewer (100) sources are used for NSBH than for BBH (200). This can be explained by reminding that for BBH we used a slightly lower SNR threshold (12 vs 13.9), thus increasing the number of weak signals that do not contribute much to the measurement, while broadening the posterior distributions. Furthermore, the characteristic effects of misaligned spins (e.g. amplitude precession) are more visible in NSBH, which make them ideal sources for this test.

We stress that we do \emph{not} assume that the priors in eq.~\ref{eq.evidence} perfectly match the corresponding distributions in the simulated events, which will likely happen in the first years of gravitational-wave astrophysics. Two important examples are the prior distributions for distance and masses.

While geometrical arguments led us to use a prior for the luminosity distance uniform in comoving volume, in reality, since far away sources would not be detectable, the distance distribution of \emph{detected} events will first increase with  distance, reach a maximum, and then decrease. The distances corresponding to the maximum of the distribution and the length of the tail depend on the true astrophysical distribution of masses (heavy CBC will be visible farther away), which we don't know (but will hopefully measure in the coming years). Similarly for the mass prior: we used priors in the component masses which were a factor of few larger than the range used to simulate the sources.
It will be interesting to verify how the test performs if the true distribution of tilt angles for the aligned model is different than what is used to split the two models, or if the true distributions are not mutually exclusive\footnote{In our simulations we assumed isotropic tilt distributions $p(\tau)\propto \sin{\tau}$, with cuts at $10^\circ$. This makes our non-aligned distribution basically equal to a fully isotropic distributions, since the probability that both tilts are small (or close to $\pi$ for BBH) is negligible.}. Given the amount of simulations that would be necessary to fully explore those scenarios, we leave it for future work.

It is worth remarking that while we consider two possible formation channels, this framework can be extended to take into account more models, provided they are mutually exclusive. Similarly, if one believes only one formation channel is possible in the universe, and thus \emph{all} events will either have aligned or not-aligned spins (that would correspond to $f_a=0,1$), model selection can be used to quantify how many detections are required before the model can be proven right. 

\begin{figure}[htb]
\includegraphics[width=0.5\textwidth]{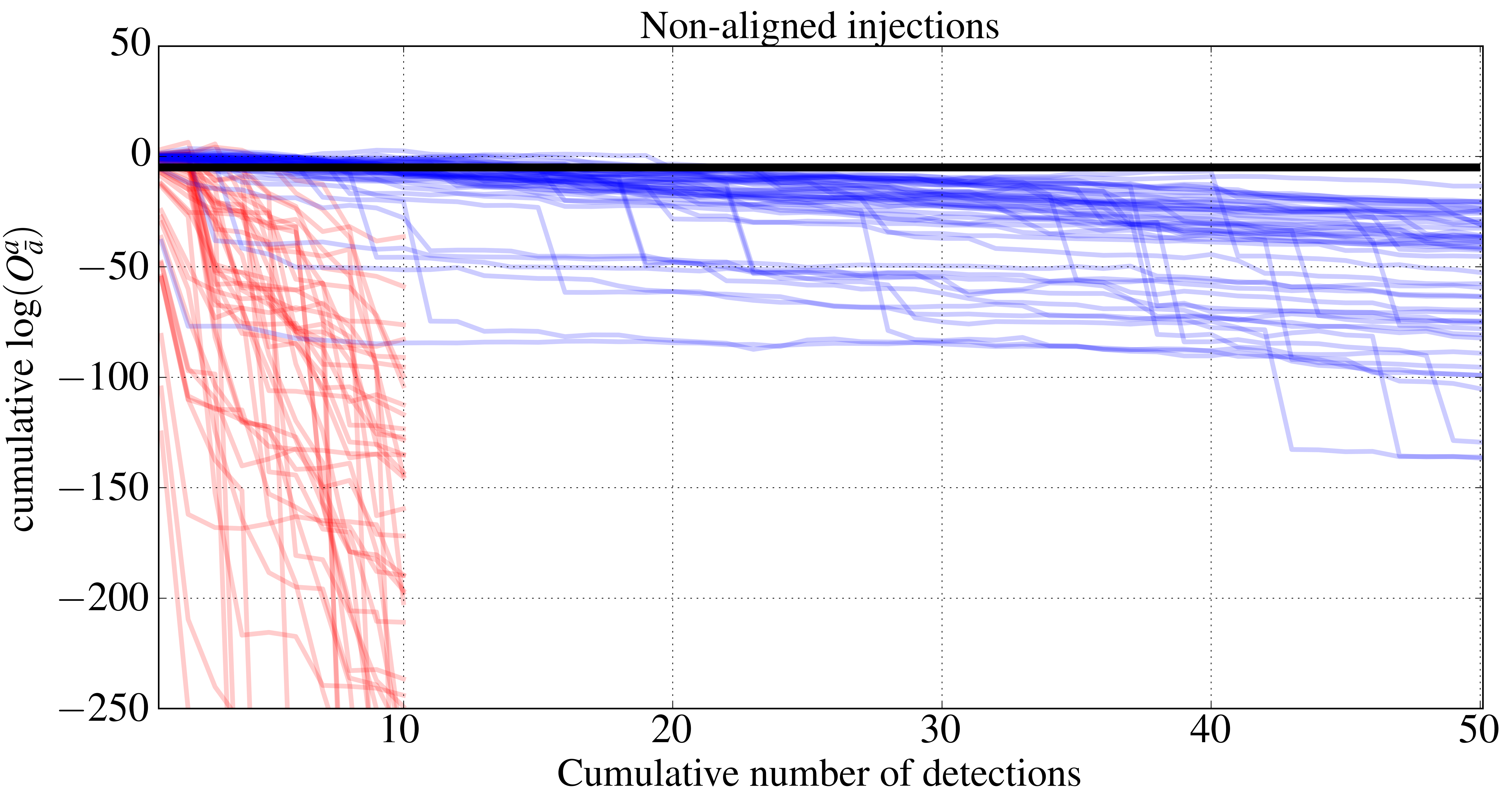}
\caption{(color online) Cumulative odds ratio for NSBH (red) and BBH (blue), with non-aligned injections. Each line is a sub-catalog. Cumulative odds values below the solid horizontal thick line favor the (correct) non-aligned model with a significance larger than \si2.7 $\sigma$. We notice that we cut the y axis at -250 to improve clarity. Some NSBH catalogs go down to cumulative odds of -1000.}
\label{Fig.Odds}
\end{figure}

Although we believe considering a mixture of two models is more consistent with today's understanding of binaries' formation, we give an example of a single-channel test. For this, we simulate a situation in which all sources have non-aligned spins and we calculate the cumulative odds ratio:
\beq
O^a_{\bar{a}}\equiv\frac{p(\mathcal{H}_a | \vec d)}{p(\mathcal{H}_{\bar{a}} | \vec{d})}= \frac{p(\vec{d}| \mathcal{H}_a) p(\mathcal{H}_a)}{p(\vec{d}| \mathcal{H}_{\bar{a}}) p(\mathcal{H}_{\bar{a}})}=B^a_{\bar{a}}~\frac{p(\mathcal{H}_a)}{p(\mathcal{H}_{\bar{a}})}\nonumber
\eeq
where $B^a_{\bar{a}}$ is the cumulative Bayes factor for aligned vs non-aligned models. Since the data corresponding to the $N$ detections is statistically independent, the cumulative Bayes factor can be written as a product over the single events:
\beq
B^a_{\bar{a}}= \prod_{k=1}^{N} \frac{p(d_k| \mathcal{H}_a)}{p(d_k| \mathcal{H}_{\bar{a}})} \equiv \prod_{k=1}^{N} \frac{Z^a_k}{Z^{\bar{a}}_k}
\eeq
The logarithm of the odds ratio is shown in Fig.~\ref{Fig.Odds} as a function of the number of events, for random sub-catalogs of 10 NSBH and 50 BBH. We have assumed $p(\mathcal{H}_a)=p(\mathcal{H}_{\bar{a}})$. We see that for both type of sources the correct non-aligned model is favored in a significant way (log odds below the solid horizontal line favor the non-aligned model at a $>2.7 \sigma$ level). NSBH curves go down faster since for NSBH the effect of spin misalignment are stronger in the waveform, and thus harder to match with an aligned model.

\section{Conclusions}\label{Sec.Conclusions}

Two formation channels are commonly considered for CBC: common envelope evolution, which should result in spins to be preferentially along (or very close to) the direction of the orbital angular momentum, and dynamical capture, which should results in randomly oriented spins. In fact, there is not complete agreement on whether common envelope evolution is efficient enough in aligning spins, or if instead eventual kicks from the core collapse supernova of the progenitor stars will be the dominant factor.
It would thus be of importance to calculate which fraction of the compact binaries have spins nearly aligned with the orbital angular momentum, which could be used to expand our understanding of formation channels. In this paper, we have shown how gravitational waves emitted by compact binaries containing a black hole, and detected by Advanced LIGO and Virgo, can be used to verify if spins are preferentially aligned with the orbital angular momentum.
We considered neutron star - black hole and binary black hole systems, and created catalogs of sources with increasingly large fraction of aligned sources (from 0 to 100\%). Black holes in NSBH were of low mass (up to 11\msun), while for BBH we simulated heavy objects, comparable to GW150914 (masses in the range $[30-50]$\msun), which will be detected in large number in the coming months and years.

We showed how a couple hundred signals are enough to pinpoint the underlying value of the aligned fraction with \si10\% uncertainty, which suggests GWs represent a viable way of gaining insight into the orientation of spins in compact binaries, and ultimately on their evolution.
We have verified the robustness of the test against some common prior mismatch (distance, masses). Future work includes introducing a mismatch between the definition of aligned in the test and the true distribution of aligned sources. We also stress that if more information is available which could help distinguish between the two channels (e.g. the resulting mass ratio distribution), it could be folded in an extended version of this test.

This is LIGO document P1500022.

\section{Acknowledgments}

SV and RL acknowledge the support of the National Science Foundation and the LIGO Laboratory. LIGO was constructed by the California Institute of Technology and Massachusetts Institute of Technology with funding from the National Science Foundation and operates under cooperative agreement PHY-0757058.
RS is supported by the FAPESP grant 2013/04538-5. 
PG was supported by NASA grant NNX12AN10G.
SV, RL and RS acknowledge the FAPESP-MIT grant 2014/50727-7.
The authors would like to acknowledge the LIGO Data Grid clusters, without which the simulations could not have been performed. Specifically, these include the Syracuse University Gravitation and Relativity cluster, which is supported by NSF awards PHY-1040231 and PHY-1104371, the Leonard E Parker Center for Gravitation, Cosmology and Astrophysics at University of Wisconsin-Milwaukee (NSF-0923409) and the Atlas cluster of the Albert Einstein Institute in Hanover.
We would also like to thank M.~Branchesi, W.~Del~Pozzo, T.~Dent, R.~Essick, W.~Farr, V.~Grinberg E.~Katsavounidis, F.~Lacasa, T.~G.~F.~Li, I.~Mandel, M.~Mapelli, C.~Van~Den Broeck, A.~Weinstein, R.~Weiss and the CBC parameter estimation group for useful comments and suggestions.
\bibliography{../pe}
\end{document}